\documentclass[twocolumn,superscriptaddress, preprintnumbers,pra]{revtex4}
\usepackage{graphicx}
\usepackage{dcolumn}
\usepackage{amssymb}
\usepackage{amsmath}
\usepackage{bm}
\usepackage{setspace}

\usepackage{hyperref}

\hypersetup
{
colorlinks=true, 
citecolor=blue
}

\begin{document}

\title{Quantitative theory of composite fermions in Bose-Fermi mixtures at $\nu=1$}
\author{Ken K. W. Ma and Kun Yang}

\affiliation{National High Magnetic Field Laboratory and Department of Physics,
Florida State University, Tallahassee, Florida 32306, USA}


\begin{abstract}
The theory of composite fermion provides a simple and unified picture to understand a vast amount of phenomenology in the quantum Hall regime. However it has remained challenging to formulate this concept properly within a single Landau level, which provides the relevant degrees of freedom in the limit of strong magnetic field. Recently
a low-energy noncommutative field theory for bosons at Landau-level filling factor $\nu=1$ has been formulated by Dong and Senthil [Z. Dong and T. Senthil, Phys. Rev. B \textbf{102}, 205126 (2020)]. In the limit of long-wavelength and small-amplitude gauge fluctuation, they found it reduces to the celebrated Halperin-Lee-Read theory of a composite fermion liquid. In this work we consider a Bose-Fermi mixture at total filling factor $\nu=1$. Different from previous work, the number density of composite fermions in the mixture and corresponding Fermi momentum can be tuned by changing the filling factor of bosons, $\nu_b = 1 -\nu_f$. This tunability enables us to study the dilute limit $\nu_b\ll 1$, which allows for a controlled and asymptotically exact calculation of the energy dispersion and effective mass of composite fermions. Furthermore, the approximation of the low-energy description by a commutative field theory is manifestly justified. Most importantly, we demonstrate that gauge fluctuations acquire a Higgs mass due to the presence of a composite boson condensate, as a result of which the system behaves like a genuine Landau Fermi liquid. Combined with the irrelevance of four-fermion interaction in the dilute limit, we are able to obtain asymptotically exact properties of this composite fermion Fermi liquid. In the opposite limit of $\nu_f\ll 1$, the Higgs mass goes to zero and we find crossover between Fermi liquid and non-Fermi liquid as temperature increases. Observing these properties either experimentally or numerically provides unambiguous evidence of not only the composite fermions and the Fermi surface they form, but also the presence of emergent gauge fields and their fluctuations due to strong correlation.

\end{abstract}

\date{\today}

\maketitle

\section{Introduction} \label{sec:intro}

The discovery of the fractional quantum Hall effect (FQHE)\cite{Tsui1982} ushered in the era of topological states of matter~\cite{Book}. Since the kinetic energy of electrons in flat Landau levels is completely quenched, the physics of FQHE is dictated by the Coulomb interaction. In studying this strongly correlated state, traditional approaches such as perturbation theory and Landau-Fermi liquid theory are powerless. An important conceptual advance was the composite fermions introduced by Jain~\cite{Jain-book}. Each composite fermion (CF) is formed by attaching an even number of magnetic flux quanta to an electron. This flux attachment leads to a reduced effective magnetic field for composite fermions. FQH states of electrons with most odd-denominator filling factors originate from filling the composite fermion Landau levels.

By incorporating the idea of flux attachment, a theory was developed by Halperin, Lee, and Read (HLR)~\cite{HLR} to describe both incompressible (or quantum Hall) and compressible states formed by CFs. In particular, when the electronic system has filling factor
$\nu=1/2$, the composite fermions experience a zero effective magnetic field under the mean-field approximation. They form a Fermi surface which couples to a fluctuating U(1) gauge field with a Chern-Simons term~\cite{footnote1}. Thus, the system is predicted to be gapless. The HLR theory has successfully explained various features of the gapless state. It also gave predictions that were verified in surface acoustic wave experiment~\cite{exp1,exp2,exp3}. However, the HLR theory is not restricted to the lowest Landau level (LLL). This restriction becomes relevant when the cyclotron gap $\hbar eB/m^*$ is much larger than the Coulomb interaction between electrons. The above conceptual issue has motivated subsequent work on formulating theories restricted to the LLL for the half-filled Landau level problem~\cite{MS1997, MS2003, MS2007, MS2016}.

Instead of studying the original problem of electrons at $\nu=1/2$, Pasquier and Haldane proposed studying a closely related but simpler problem, namely the system of bosons at $\nu=1$~\cite{Pasquier}. By attaching a \textit{single} magnetic flux quantum to each boson, the bosons become CFs which experience a zero average effective magnetic field, just as electrons at $\nu=1/2$. Then, the theory for bosons restricted to the LLL was constructed in an enlarged Hilbert space of composite fermions. In addition, a set of constraints were posed so that the original Hilbert space can be recovered. This approach was further developed by Read, who used it to derive response functions of the system~\cite{Read1998}.

Motivated by subsequent developments of noncommutative field theory in high-energy and in particular, string theory~\cite{Seiberg-Witten, Douglas-RMP, Szabo}, Dong and Senthil reinitiated the studies of the $\nu=1$ boson problem after two decades~\cite{Senthil2020}. They have successfully formulated a description of the system in terms of a noncommutative field theory. Furthermore, they apply the Seiberg-Witten map~\cite{Seiberg-Witten} to the resultant theory, and showed it maps onto the HLR theory in the lowest order approximation. In making this approximation, the parameter $|\Theta|$ in the commutator, $[R_x,R_y]=i\Theta=-i\ell_B^2$ where $R_x$ and $R_y$ are guiding center coordinates, plays the important role of expansion parameter. Here, $\ell_B=\sqrt{1/eB}$ is the magnetic length of the system in units of $\hbar=c=1$. Since $\Theta$ is dimensionful, it needs to combine with other physical quantities to form dimensionless parameters. Hence, the approximation by the HLR theory is justified in the long-wavelength regime and the fluctuation in gauge field has a small amplitude. The above approach has been applied in a series of recent work in studying quantum Hall physics and related topics~\cite{Milovanovic2021, Senthil2021, Goldman, Son2021}.

Despite the success and continuous progress in previous work, it remains difficult to study properties of the CFs quantitatively, like their effective mass. Furthermore the presence of strong gauge fluctuations renders the properties of the CF liquids hard to access reliably, even in a qualitative manner. To address and provide deeper insights into the above conceptual challenges, we study a different variant of the original problem of electrons at $\nu=1/2$, namely Bose-Fermi mixtures at total filling factor $\nu=\nu_f+\nu_b = 1$, where $\nu_f$ and $\nu_b$ are the fermion and boson filling factors respectively. Among other things we obtain an asymptotically exact solution in the limit $\nu_b\rightarrow 0$, where the system behaves as a gas of {\em free} CFs. Importantly, Bose-Fermi mixtures have been realized in cold atom systems~\cite{McNamara2006, FB2014, FB2015, Chin2017, Yao2016, Roy2017, Wu2018, footnote}. Different from two-dimensional electronic systems, the formation of Landau levels and realization of quantum Hall physics in cold atoms would require a rapid rotation~\cite{Cornell, Viefers, Madison, Ketterle} or a synthetic gauge field~\cite{Spielman, Spielman2} in the system. Although reaching the quantum Hall regime in cold atom systems still remains a great challenge, the continuous progress in fabricating synthetic gauge field in optical lattices makes the direction increasingly promising. Thus, the system we consider here is as experimentally accessible (albeit challenging) as the system of bosons at $\nu=1$, and can be tested experimentally in principle. In the following, we first demonstrate the unique properties of the system using a simple argument based on flux attachment. 

By attaching a single magnetic flux quantum to every particle in the mixture, the original bosons and fermions are turned into composite fermions and composite bosons (CBs) respectively, both experiencing zero effective magnetic field. In contrast to previous work, the number density of composite fermions and the corresponding size of the Fermi surface can be tuned by modifying the filling factor of bosons $\nu_b$ (and that of fermions, $\nu_f$  accordingly). The existence of CBs significantly modifies the effect of gauge field fluctuation in the mixture. Due to the Bose-Einstein condensation (BEC) of CBs, the gauge field acquires a Higgs mass, rendering their effects on the Fermi surface of CFs weak; as a result the CFs survive as genuine Landau quasiparticles at low energy. Furthermore small $\nu_b$ suppresses the momentum of the gauge bosons that couples to CFs near the Fermi surface due to the small CF Fermi surface, which justifies the application of Seiberg-Witten map to the lowest order of $k_F \ell_B$ where $k_F$ is the Fermi momentum. Another advantage of this limit is the CF dispersion can be obtained exactly in this limit, and the interaction between composite fermions are irrelevant. The combination of these allow for an asymptotically exact solution in this limit, including the exact coefficient of the linear $T$ specific heat.

\begin{figure} [htb]
\begin{center}
\includegraphics[width=3.3in]{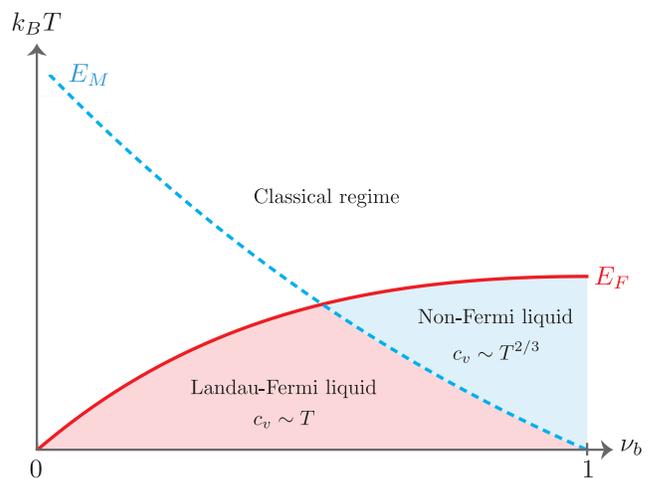}
\caption{A generic illustration of the crossover from Landau-Fermi liquid to non-Fermi liquid for composite fermions in the Bose-Fermi mixture at total filling factor $\nu=1$. See Sec.~\ref{sec:crossover} for a discussion of the energy scales $E_F$ and $E_M$.}
\label{fig:crossover}
\end{center}
\end{figure}

As we will show later, the mass term of the gauge field is controlled by the condensate density of BEC, which is proportional to the filling factor of composite bosons (or equivalently, original fermions). This points to a crossover between a Landau-Fermi liquid and a non-Fermi liquid regimes in the system. The crossover can be probed by the change in temperature dependence of specific heat capacity at low temperature. Our results are summarized in Fig. \ref{fig:crossover}. Observing the behavior of Fig. \ref{fig:crossover} provides unambiguous evidence of not only the composite fermions and the Fermi surface they form, but also the presence of emergent gauge fields and their fluctuations due to strong correlation.

\section{The Pasquier-Haldane-Read construction}  \label{sec:BF}

Now, we describe the two-dimensional Bose-Fermi mixture in more detail. The system consists of $N_b$ bosons and $N_f$ fermions, which is placed under a transverse magnetic field (either real or artificial depending on the actual system). We assume the boson and fermion have the same charge. The lowest Landau level has a degeneracy $N=N_b+N_f$, so that the system has a total filling factor $\nu=1$. The corresponding filling factors for bosons and fermions are $\nu_b=N_b/N$ and $\nu_f=N_f/N=1-\nu_b$.

The \textit{physical} Hilbert space of the system is spanned by the following set of unnormalized $N$-particle states,
\begin{eqnarray} \label{eq:phy-state}
|\Psi_{N_b, N_f}\rangle
=a^\dagger_{m_1}\cdots a^\dagger_{m_{N_b}}
f^\dagger_{s_1}\cdots f^\dagger_{s_{N_f}}
|\rm{vac}\rangle.
\end{eqnarray}
Here, $|\text{vac}\rangle$ denotes the vacuum state with no particles. The operators $a^\dagger_m$ and $f^\dagger_m$ create a boson and a fermion in the $m$-th orbital in the LLL, respectively. Both indices $m_i$ and $s_i$ can take value from $1$ to $N$. As a result, there are $\binom{N_b+N-1}{N_b}\times\binom{N}{N_f}$ linearly independent basis states in the Hilbert space.

To formulate an effective theory restricted to the lowest Landau level, various density operators in the system should be projected to the LLL. Each of the projected density operator in the momentum space $\rho(\mathbf{q})$ satisfies the Girvin-MacDonald-Platzman (GMP) algebra~\cite{GMP},
\begin{eqnarray} \label{eq:GMP}
\left[\rho(\mathbf{q}), \rho(\mathbf{p})\right]
=2i\sin{\left(\frac{\mathbf{q}\times\mathbf{p}}{2}\ell_B^2\right)}
\rho\left(\mathbf{q}+\mathbf{p}\right).
\end{eqnarray}
Hence, it is necessary to find a representation of density operators which satisfy the GMP algebra. A possible solution was proposed by Pasquier and Haldane~\cite{Pasquier} and further developed by Read~\cite{Read1998}, as we review below.

\subsection{A brief review of bosons at $\nu=1$}
\label{sec:boson}

To be comprehensive, we first summarize briefly the construction for bosons at $\nu=1$ by Pasquier and Haldane~\cite{Pasquier}. This review will turn out useful in our later construction for the Bose-Fermi mixture. Our following discussion follows closely the presentation in Ref.~\cite{Read1998}. The physical Hilbert space of $N$ bosons is spanned by the basis states
\begin{eqnarray}
|\Psi_b\rangle
=a_{m_1}^\dagger\cdots a_{m_N}^\dagger |\text{vac}\rangle.
\end{eqnarray}

Instead of formulating the theory in the original Hilbert space of bosons, the Pasquier-Haldane construction employs the \textit{enlarged} Hilbert space of composite fermions to represent the GMP algebra. Each CF can be viewed as a bound state of a \textit{physical} boson and a fermionic vortex. The corresponding CF operators $c_{mn}$ and $c_{nm}^\dagger$ have two indices. These operators satisfy the following anticommutation relations,
\begin{align}
\{c_{mn}, c_{n'm'}^\dagger\}
&=\delta_{mm'}\delta_{nn'},
\\
\{c_{mn}, c_{m'n'}\}
&=0,
\\ \label{eq:cdag-cdag}
\{c_{nm}^\dagger, c_{n'm'}^\dagger\}
&=0.
\end{align}
Following the convention in Refs.~\cite{Read1998, Senthil2020}, we call $m$ the left index and $n$ as the right index. The indices $m$ and $n$ label the single particle states of physical bosons and vortices, respectively. Both indices take value from $1$ to $N$, where $N$ is the degeneracy of the Landau level. From the CF operators, one defines the left density operator $\rho^{F,L}_{mm'}$ and the right density operator $\rho^{F,R}_{nn'}$ as
\begin{align}
\label{eq:FLmm}
\rho^{F,L}_{mm'}=\sum_n c^\dagger_{nm'}c_{mn},
\\
\label{eq:FRnn}
\rho^{F,R}_{nn'}=\sum_m c^\dagger_{n'm}c_{mn}.
\end{align}
It is important to note that we have added a superscript $F$ to state clearly the density operators are for composite fermions. This notation will be particularly useful in later discussion. Mathematically, $\rho^{F,L}_{mm'}$ and $\rho^{F,R}_{nn'}$ are the $N^2$ generators of U$(N)$ groups in the left and right indices, respectively. These are symmetry operations originating from the freedom in the change of basis in the Landau orbitals for the physical bosons and vortices.

The enlarged Hilbert space of $N$ CFs is spanned by the following
$\binom{N^2}{N}$ basis states:
\begin{eqnarray}
|\Psi_{\rm CF}\rangle
=c_{n_1m_1}^\dagger \cdots c_{n_N m_N}^\dagger|\text{vac}\rangle.
\end{eqnarray}
The physical Hilbert space of bosons is the subspace being spanned by the states,
\begin{eqnarray}
|\Psi_{\rm phys}\rangle
=\sum_{n_1, \cdots, n_N}
\epsilon^{n_1 \cdots n_N}
c_{n_1m_1}^\dagger \cdots c_{n_N m_N}^\dagger|\text{vac}\rangle.
\end{eqnarray}
Here, the symbol $\epsilon^{n_1\cdots n_N}$ denotes the Levi-Civita tensor. It is clear that
$|\Psi_{\rm phys}\rangle$ transforms as a singlet under the SU$(N)_R$ group:
\begin{eqnarray} \label{eq:con-boson}
\rho^{F,R}_{nn'}|\Psi_{\rm phys}\rangle
=\delta_{nn'}|\Psi_{\rm phys}\rangle.
\end{eqnarray}
In addition, consider the U$(1)$ generator given by
\begin{eqnarray}
\hat{N}
=\text{Tr}\left(\rho^{F,L}_{mm'}\right)
=\text{Tr}\left(\rho^{F,R}_{nn'}\right).
\end{eqnarray}
The physical state satisfies $\hat{N}|\Psi_{\rm phys}\rangle=N|\Psi_{\rm phys}\rangle$, which fixes the system to have $N$ physical bosons and composite fermions. From Eq.~\eqref{eq:cdag-cdag}, one can show that $|\Psi_{\rm phys}\rangle$ is symmetric under the exchange in the index $m$. Therefore, there are $2N-1\choose N$ independent states in the projected Hilbert space. Both symmetry condition and dimension agree with the physical Hilbert space of $N$ bosons.

To verify the density operators defined in the enlarged Hilbert space do satisfy the GMP algebra, it is necessary to represent operators in the momentum space. The CF operators in the momentum space representation $c_{\mathbf{q}}$ is related to $c_{mn}$ by
\begin{eqnarray} \label{eq:cq}
c_{mn}
=\int\frac{d^2\mathbf{q}}{(2\pi)^{3/2}}
\langle m|e^{i\mathbf{q}\cdot\mathbf{R}}|n\rangle
c_{\mathbf{q}}.
\end{eqnarray}
Note that the symbol $\mathbf{R}=R_x\hat{x}+R_y\hat{y}$ denotes the guiding center coordinates of a particle in a magnetic field, and $e^{i\mathbf{q}\cdot\mathbf{R}}$ is proportional to a plane wave factor $e^{i\mathbf{q}\cdot\mathbf{r}}$ projected to the LLL. The number of independent ${\bf q}$'s is precisely $N^2$ in a system with geometry of a torus (so that it is translationally invariant), exactly matching the number of $(mn)$ indices. In addition, the normalization factor
$1/(2\pi)^{3/2}$ follows the convention in Ref.~\cite{Read1998}. Based on this normalization, the operator $c_{\mathbf{q}}$ satisfies the anticommutation relation,
\begin{eqnarray} \label{eq:anticommutator}
\left\{c_\mathbf{q}, c_\mathbf{p}^\dagger\right\}
=(2\pi)^2\delta(\mathbf{q}-\mathbf{p}).
\end{eqnarray}
The left density and right density operators in the momentum space can be obtained from
$c_\mathbf{q}$ and $c_\mathbf{q}^\dagger$ as
\begin{align}
\label{eq:F-L}
\rho_{F, L}(\mathbf{q})
&=\int \frac{d^2\mathbf{k}}{(2\pi)^2}c_{\mathbf{k}-\mathbf{q}}^\dagger c_{\mathbf{k}}
~e^{i(\mathbf{k}\times\mathbf{q})\ell_B^2/2},
\\
\label{eq:F-R}
\rho_{F, R}(\mathbf{q})
&=\int \frac{d^2\mathbf{k}}{(2\pi)^2}c_{\mathbf{k}-\mathbf{q}}^\dagger c_{\mathbf{k}}
~e^{-i(\mathbf{k}\times\mathbf{q})\ell_B^2/2}.
\end{align}

Using Eq.~\eqref{eq:anticommutator}, it is straightforward to show that $\rho_{F, L}(\mathbf{q})$ indeed satisfies the GMP algebra in Eq.~\eqref{eq:GMP}. One can also show
\begin{eqnarray}
\left[\rho_{F,L}(\mathbf{q}), \rho_{F,R}(\mathbf{p})\right]=0,
\end{eqnarray}
which implies the SU$(N)$ symmetries in the left and right indices are separate. This fact will be important when we couple the composite fermions to gauge fields. The constraint to the right density operator [Eq.~\eqref{eq:con-boson}] becomes
\begin{eqnarray}
\rho_{F, R}(\mathbf{q})
=(2\pi)^2\bar{\rho}\delta(\mathbf{q}),
\end{eqnarray}
where $\bar{\rho}=1/(2\pi\ell_B^2)$ is the average density of particles at $\nu=1$.

\subsection{Bose-Fermi mixture at $\nu=1$} \label{sec:mixture}

The above review has set the stage for us to apply the Pasquier-Haldane construction in the Bose-Fermi mixture at $\nu=1$. As suggested by the simple argument of flux attachment in Sec.~\ref{sec:intro}, we need to include both CFs and CBs in the enlarged Hilbert space. We use $b_{mn}$ and $b_{nm}^\dagger$ to denote the annihilation and creation operators for composite bosons, respectively. These operators satisfy
\begin{align}
\label{eq:CC-boson}
[b_{mn}, b^\dagger_{n'm'}]
&=\delta_{mm'}\delta_{nn'},
\\
[b_{mn}, b_{m'n'}]
&=0,
\\
[b^\dagger_{nm}, b^\dagger_{n'm'}]
&=0.
\end{align}
For composite fermions, they are described by the same set of operations in Sec.~\ref{sec:boson}. Similar to Eqs.~\eqref{eq:FLmm} and~\eqref{eq:FRnn}, the associated left and right density operators for composite bosons are
\begin{align}
\rho^{B,L}_{mm'}=\sum_n b^\dagger_{nm'}b_{mn},
\\
\rho^{B,R}_{nn'}=\sum_m b^\dagger_{n'm}b_{mn}.
\end{align}

For the Bose-Fermi mixture, the physical state $|\Psi_{\rm Phys}^{N_b, N_f}\rangle$ in Eq.~\eqref{eq:phy-state} can be obtained from
\begin{align} \label{eq:general-state}
\nonumber
|\Psi_{\rm Phys}^{N_b, N_f}\rangle
=\sum_{\substack{n_1\cdots n_{N_b} \\ j_1\cdots j_{N_f}}}
&\epsilon^{n_1\cdots n_{N_b} j_1\cdots j_{N_f}}
\left[c_{n_1m_1}^\dagger \cdots c_{n_{N_b} m_{N_b}}^\dagger
\right.
\\
& \times\left.
 b_{j_1s_1}^\dagger \cdots b_{j_{N_f} s_{N_f}}^\dagger\right]
|0\rangle.
\end{align}
Here, all indices $n_i, m_i, j_i, s_i$ take value from $1$ to $N$. Clearly, the state in Eq.~\eqref{eq:general-state} is symmetric under the exchange in index $m$ and antisymmetric under the exchange in index $s$. Furthermore, it has no effect under the exchange between the $m$ and $s$ indices. All these symmetries are consistent with the many-body states in Eq.~\eqref{eq:phy-state}. Now, one may wonder why the antisymmetrization is carried over both indices $n$ and $j$, but not antisymmetrizing each of them separately. There are two reasons. First, it ensures that both $c_{nm}^\dagger$ and $b_{js}^\dagger$ are square matrices, with the Levi-Civita tensor being well-defined. Another more physical reason is that both bosons and fermions are coupled to the same set of vortices. Therefore, the SU$(N)_R$ symmetry should remain the same as in the system of bosons at $\nu=1$. Also,
$|\Psi_{\rm Phys}^{N_b, N_f}\rangle$ should transform as a singlet under the SU$(N)_R$ group.

\begin{figure} [htb]
\begin{center}
\includegraphics[width=2.0in]{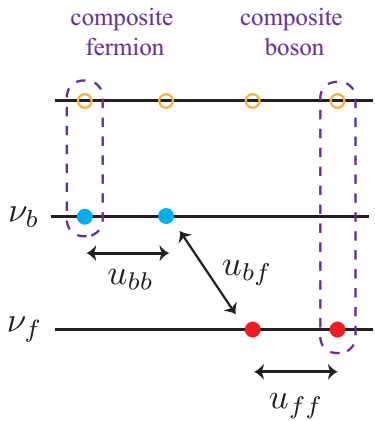}
\caption{The Pasquier-Haldane construction in the bilayer system at total filling factor $\nu=1$. Different types of interactions $u_{bb}$, $u_{ff}$, and $u_{bf}$ are defined in Sec.~\ref{sec:H-BF}. Here, the blue solid dots, red solid dots, and the yellow empty circles denote the physical bosons, physical fermions, and the vortices, respectively. The dashed lines illustrate composite particles. Each of them is formed between a physical particle and a vortex.}
\label{fig:PHR}
\end{center}
\end{figure}

The physical state in Eq.~\eqref{eq:general-state} satisfies the following condition,
\begin{eqnarray} \label{eq:constraint-nn}
(\rho^{F, R}_{nn'}+\rho^{B, R}_{nn'})
|\Psi_{\rm Phys}^{N_b, N_f}\rangle
=\delta_{nn'}|\Psi_{\rm Phys}^{N_b, N_f}\rangle.
\end{eqnarray}
In addition, the state also satisfies the following global U$(1)$ conservation laws
\begin{align}
\hat{N}_{F,L}|\Psi_{\rm Phys}^{N_b, N_f}\rangle
&=\sum_{m=1}^{N}\rho^{F,L}_{mm}|\Psi_{\rm Phys}^{N_b, N_f}\rangle
=N_b|\Psi_{\rm Phys}^{N_b, N_f}\rangle,
\\
\hat{N}_{F,R}|\Psi_{\rm Phys}^{N_b, N_f}\rangle
&=\sum_{n=1}^N\rho^{F,R}_{nn}|\Psi_{\rm Phys}^{N_b, N_f}\rangle
=N_b|\Psi_{\rm Phys}^{N_b, N_f}\rangle,
\\
\hat{N}_{B,L}|\Psi_{\rm Phys}^{N_b, N_f}\rangle
&=\sum_{s=1}^N\rho^{B,L}_{ss}|\Psi_{\rm Phys}^{N_b, N_f}\rangle
=N_f|\Psi_{\rm Phys}^{N_b, N_f}\rangle,
\\
\hat{N}_{B,R}|\Psi_{\rm Phys}^{N_b, N_f}\rangle
&=\sum_{j=1}^N\rho^{B,R}_{jj}|\Psi_{\rm Phys}^{N_b, N_f}\rangle
=N_f|\Psi_{\rm Phys}^{N_b, N_f}\rangle.
\end{align}
The above conservation laws simply indicate that there are $N_b$ CFs and $N_f$ CBs in the system. These numbers are equal to the numbers of physical bosons and fermions, respectively. Thus, the filling factor of CFs is $\nu_b$, which \textit{can be tuned} by changing $N_b$ and $N_f$ accordingly! The above discussion is summarized in Fig.~\ref{fig:PHR}.

In momentum space, the composite boson operators can be constructed analogous to Eq. (\ref{eq:cq}) and satisfy the commutation relation
\begin{eqnarray} \label{eq:commutator}
\left[b_\mathbf{q}, b_\mathbf{p}^\dagger\right]
=(2\pi)^2\delta(\mathbf{q}-\mathbf{p}),
\end{eqnarray}
which is analogous to the anti-commutation relation Eq.~\eqref{eq:anticommutator} for CF operators.
The density operators for composite fermions are still given by Eqs.~\eqref{eq:F-L} and~\eqref{eq:F-R}. For composite bosons, the left and right density operators in momentum space are
\begin{align}
\label{eq:B-L}
\rho_{B, L}(\mathbf{q})
&=\int \frac{d^2\mathbf{k}}{(2\pi)^2}b_{\mathbf{k}-\mathbf{q}}^\dagger b_{\mathbf{k}}
~e^{i(\mathbf{k}\times\mathbf{q})\ell_B^2/2},
\\ \label{eq:B-R}
\rho_{B, R}(\mathbf{q})
&=\int \frac{d^2\mathbf{k}}{(2\pi)^2}b_{\mathbf{k}-\mathbf{q}}^\dagger b_{\mathbf{k}}
~e^{-i(\mathbf{k}\times\mathbf{q})\ell_B^2/2}.
\end{align}
It is straightforward to verify that $\rho_{F, L}(\mathbf{q})$ and $\rho_{B, L}(\mathbf{q})$ satisfy the GMP algebra separately,
\begin{align}
\nonumber
\left[\rho_{F,L}(\mathbf{q}), \rho_{F,L}(\mathbf{p})\right]
&=2i\sin{\left(\frac{\mathbf{q}\times\mathbf{p}}{2}\ell_B^2\right)}
\rho_{F,L}(\mathbf{q}+\mathbf{p}),
\\ \nonumber
\left[\rho_{B,L}(\mathbf{q}), \rho_{B,L}(\mathbf{p})\right]
&=2i\sin{\left(\frac{\mathbf{q}\times\mathbf{p}}{2}\ell_B^2\right)}
\rho_{B,L}(\mathbf{q}+\mathbf{p}),
\\
\left[\rho_{F,L}(\mathbf{q}), \rho_{B,L}(\mathbf{p})\right]
&=0.
\end{align}
Finally, the constraint in Eq.~\eqref{eq:constraint-nn} is translated into
\begin{eqnarray} \label{eq:constraint}
G(\mathbf{q})
=\rho_{F,R}(\mathbf{q})+\rho_{B,R}(\mathbf{q})
=(2\pi)^2\bar{\rho}\delta(\mathbf{q}).
\end{eqnarray}
Since the $N$ vortices are shared among $N=N_b+N_f$ particles, $\bar{\rho}=1/(2\pi\ell_B^2)$ matches the average density of vortices.

\subsection{Composite fermions as dipolar excitons} \label{sec:dipole-pic}

To illustrate the above Pasquier-Haldane construction in a more transparent manner, we consider various simple cases here. Along the way, we will also relate the composite fermions in the mixture to the dipolar excitons in bilayer systems studied in previous work~\cite{Yang2001}. By doing so, one can appreciate the value of the construction better, and understand the physical meaning of composite fermions and composite bosons in the system.

To begin, consider the special case when all $N$ particles in the system are fermions. This is the integer quantum Hall state at $\nu=1$. Since there is only one possible way for filling $N$ fermions in $N$ different Landau orbitals in the LLL, the Hilbert space is one-dimensional. Meanwhile, there are $N$ composite bosons in the Pasquier-Haldane construction. It seems that there are $N^2$ different momenta $\mathbf{q}$ for each CB to take, similar to a CF as suggested in Eq.~\eqref{eq:cq}. This leads to a Hilbert space with a large dimension of $\binom{N^2+N-1}{N}$. However, due to the constraint (\ref{eq:constraint-nn}), the CBs actually form the unique uniform state with one particle per Landau oribtal. It is instructive to compare with the flux-attachment treatment of Zhang, Hansson, and Kivelson ~\cite{ZHK1989}, in which the CBs condense into the $\mathbf{q}=\mathbf{0}$ state to minimize the energy of the system. All excited states at $\nu=1$ involve higher Landau levels; as a result the CB BEC ground state is the only state left in the Pasquier-Haldane construction!

\begin{figure} [htb]
\begin{center}
\includegraphics[width=2.5in]{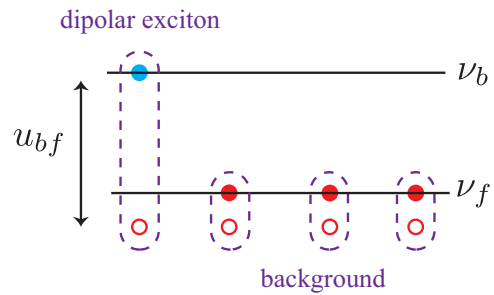}
\caption{The identification of dipolar excitons in the Bose-Fermi mixture as composite fermions. Here, the vortices are interpreted as the holes of physical fermions (marked as hollow circles). The blue and red solid dots denote physical bosons and physical fermions, respectively.}
\label{fig:dipole}
\end{center}
\end{figure}

We now consider the more interesting case of one boson and $N-1$ fermions. Equivalently, we can think of the system as being made of one boson and one fermionic hole (with opposite charge of the boson), with a Hilbert space of dimension $N\times N = N^2$. The situation here is identical to the formation of bosonic excitons from binding electrons and holes in the bilayer system at total filling factor $\nu=1$~\cite{Yang2001}, in the extreme case of one electron in the upper layer and $N-1$ electrons (or equivalently, one hole) in the lower layer. The electron-hole pair form a dipolar exciton with $N^2$ possible values of ${\bf q}$, just like a single spin-wave excitation in the $\nu=1$ quantum Hall ferromagnet~\cite{Kallin1984}. The difference in statistics for the single marked particle (boson in the present case and electron in the upper layer in Ref.~\cite{Yang2001}) has no effect as there are no other identical particles that it can exchange with.
In the Pasquier-Haldane construction, we have one CF and $N-1$ CBs. With the CBs forming a BEC and drop out as in the previous case, we are left with one CF that can occupy one of the $N^2$ plane wave states with fixed ${\bf q}$, matching the Hilbert space of dimension $N^2$. This immediately suggests physically a CF should be viewed as a dipole formed by the boson and fermioic hole, as illustrated in Fig.~\ref{fig:dipole}. The (single) CF dispersion is identical to that of Refs.~\cite{Kallin1984, Yang2001}, and we will demonstrate how to obtain it systematically from the Pasquier-Haldane construction in Sec.~\ref{sec:single}.


The situation becomes more complicated when the mixture has two or more bosons. Consider the specific case with two bosons, or equivalently, two composite fermions. Naively there are $N^2(N^2-1)/2$ ways to put two CFs in the $N^2$ CF momentum eigenstates, bigger than the actual physical Hilbert space size of $N^2(N^2-1)/4$ by a factor of two, similar to what happens in the bilayer system of electrons. As pointed out in Ref.~\cite{Yang2001}, this factor of two comes from the ambiguity in associating one boson (in the present case) with one of the two fermioic holes. Due to this overcompleteness one cannot treat each dipolar exciton there as a point particle. The Pasquier-Haldane construction provides a systematic recipe to resolve the overcompleteness problem by imposing a singlet representation of the SU$(N)_R$ group for composite particles. This also leads to the coupling between composite particles and an emergent gauge field, which will be discussed in Sec.~\ref{sec:gauge}.

\section{Hamiltonian of the Bose-Fermi mixture} \label{sec:H-BF}

Armed with the machinery and physical picture developed in the previous section, we are now ready to formulate the Hamiltonian for the Bose-Fermi mixture. Our main goal is to construct a suitable Hamiltonian from which the exact single-particle dispersion emerges, and the two-body scattering matrix elements of composite fermions match previously known results. Also, the constraint in Eq.~\eqref{eq:constraint} is satisfied in the construction.

Starting from the physical Hilbert space, the Hamiltonian describing the interaction between particles in the Bose-Fermi mixture is given by
\begin{align}
\nonumber
H_0
=~&\frac{1}{2}\sum_{m_1,m_2,m_3,m_4}
(u_{bb})_{m_1,m_2,m_3,m_4}
a^\dagger_{m_1}a^\dagger_{m_2}a_{m_4}a_{m_3}
\\ \nonumber
&+\frac{1}{2}\sum_{m_1,m_2,m_3,m_4}
(u_{ff})_{m_1,m_2,m_3,m_4}
f^\dagger_{m_1}f^\dagger_{m_2}f_{m_4}f_{m_3}
\\
&+\sum_{m_1,m_2,m_3,m_4}
(u_{bf})_{m_1,m_2,m_3,m_4}
a^\dagger_{m_1}f^\dagger_{m_2}f_{m_4}a_{m_3}.
\end{align}
The symbol $(u_{bb})_{m_1,m_2,m_3,m_4}$ denotes the matrix element of the boson-boson interaction projected to the lowest Landau level~\cite{Read1998}:
\begin{widetext}
\begin{eqnarray}
(u_{bb})_{m_1,m_2,m_3,m_4}
=\int
u_{bb}(\mathbf{x}_1-\mathbf{x}_2)
\overline{\varphi_{m_1}(z_1)}\overline{\varphi_{m_2}(z_2)}
\varphi_{m_3}(z_1)\varphi_{m_4}(z_2)
~d^2 \mathbf{x}_1 d^2 \mathbf{x}_2.
\end{eqnarray}
\end{widetext}
Here, $\varphi_m(z)$ is the single particle basis wave function in the LLL. We use the overbar to denote complex conjugation. Similar definition holds for the matrix elements of $u_{ff}$ and $u_{bf}$. These two terms describe the fermion-fermion interaction and fermion-boson interaction, respectively.

In order to represent the Hamiltonian in terms of the CFs and CBs introduced earlier, we employ the density operators defined in Eqs.~\eqref{eq:F-L}, \eqref{eq:F-R}, \eqref{eq:B-L}, and~\eqref{eq:B-R} which satisfy the GMP algebra. The corresponding Hamiltonian is
\begin{widetext}
\begin{align}  \label{eq:H0}
\nonumber
H
=~&\frac{1}{2}\int \frac{d^2q}{(2\pi)^2}
U_{bb}(\mathbf{q}):\rho_{F,L}(\mathbf{q})\rho_{F,L}(\mathbf{-q}):
+\frac{1}{2}\int \frac{d^2\mathbf{q}}{(2\pi)^2}
U_{ff}(\mathbf{q}):\rho_{B,L}(\mathbf{q})\rho_{B,L}(\mathbf{-q}):
\\
&+\frac{1}{2}\int \frac{d^2\mathbf{q}}{(2\pi)^2}
U_{bf}(\mathbf{q})
:\left[
\rho_{B,L}(\mathbf{q})\rho_{F,L}(\mathbf{-q}):
+\rho_{F,L}(\mathbf{q})\rho_{B,L}(\mathbf{-q})\right]:
\end{align}
\end{widetext}
where
\begin{align}
U(\mathbf{q})
=\tilde{u}(\mathbf{q})e^{-q^2\ell_B^2/2}
=e^{-q^2\ell_B^2/2}\int u(\mathbf{x}) e^{-i\mathbf{q}\cdot\mathbf{x}}~d^2x.
\end{align}
The notation $:\mathcal{O}:$ denotes the normal ordering of operator $\mathcal{O}$. Since
$\rho_L^\dagger(\mathbf{q})=\rho_L(-\mathbf{q})$, we intentionally write the second line in the present form to show the Hermiticity explicitly. The Gaussian factor in $U(\mathbf{q})$ originates from the definition of LLL-projected density operator~\cite{Read1998}. Furthermore, Eq.~\eqref{eq:H0} can be written in the form,
\begin{widetext}
\begin{align} \label{eq:H0-matrix}
\nonumber
H
=~&\frac{1}{2}\sum_{m_1,m_2,m_3,m_4}\sum_{n_1,n_2}~
(u_{bb})_{m_1,m_2,m_3,m_4}
~c^\dagger_{n_1,m_1}c^\dagger_{n_2, m_2}c_{m_4, n_2}c_{m_3, n_1}
\\ \nonumber
&+\frac{1}{2}\sum_{m_1,m_2,m_3,m_4}\sum_{n_1,n_2}~
(u_{ff})_{m_1,m_2,m_3,m_4}
~b^\dagger_{n_1,m_1}b^\dagger_{n_2,m_2}b_{m_4,n_2}b_{m_3,n_1}
\\
&+\sum_{m_1,m_2,m_3,m_4}\sum_{n_1,n_2}~
(u_{bf})_{m_1,m_2,m_3,m_4}
~c^\dagger_{n_1,m_1}b^\dagger_{n_2,m_2}b_{m_4, n_2}c_{m_3, n_1},
\end{align}
\end{widetext}
which commutes with the constraint in Eq.~\eqref{eq:constraint-nn}. Thus, $H$ does describe the particles in the physical Hilbert space.

\subsection{The preferred Hamiltonian in the enlarged Hilbert space}  \label{sec:pre-H}

Since we are now working in the enlarged Hilbert space, there are some freedoms in representing the physical Hamiltonian, as long as the representation is faithful in the physical Hilbert space, and there is no coupling between physical and unphysical degrees of freedom. It is therefore desirable to find the so-called ``\textit{preferred Hamiltonian}"~\cite{MS2003}, that is the form being the easiest to use when making approximations, and yield the most accurate results. To do that, we need insights on the actual physics of the system, which is provided from the picture of dipolar excitons in Sec.~\ref{sec:dipole-pic} and Fig.~\ref{fig:dipole}.

We start from the Hamiltonian in Eq.~\eqref{eq:H0}. Due to the normal ordering in all three terms, there is no one-body term which gives rise to the single-particle dispersion. This issue is commonly resolved by removing the normal ordering~\cite{Senthil2020}. In the present case, we know that a single CF at momentum $\mathbf{k}$ must have its dispersion solely depending on $u_{bf}$, because it is made of a boson and a fermionic hole (similar to the fact that the single exciton dispersion depends solely on in inter-layer interaction in Ref. \cite{Yang2001}). This observation suggests removing the normal ordering in the last term in Eq.~\eqref{eq:H0}. Furthermore, the composite bosons should also have a single-particle dispersion (which is obvious if one considers the extreme case with $\nu_b=0$ where the ground state corresponds to a BEC). This suggests removing the normal ordering in the second term in Eq.~\eqref{eq:H0} as well.

After removing the normal ordering, the Hamiltonian still cannot lead to the correct dispersion of a single composite fermion. Here, we recall a technique proposed by Murthy and Shankar~\cite{MS1997}, which has been commonly used in subsequent work~\cite{Murthy1999, Shankar2001, MS2003, MS2007, MS2016, Senthil2020}. Recall that the physical states satisfy Eq.~\eqref{eq:constraint}, and the Hamiltonian satisfies $[H, G(\mathbf{q})]=0$. In the usual case of having a single type of composite particles, one may include the constraint by replacing the density operator as $\rho_L(\mathbf{q})-G(\mathbf{q})$. This is viewed as a short cut of the conserving approximation employed by Read in Ref.~\cite{Read1998}.

Now, the mixture consists of two different types of composite particles. Hence, we still have the ambiguity in replacing which density operators. Meanwhile, the following intuition may help. In the usual case, the vortices somehow play a role similar to the holes of the physical particles. The vortex has an opposite charge of the physical particle. The binding between them leads to a neutral dipole. Then, the left density operator is replaced by $\rho_L(\mathbf{q})-\rho_R(\mathbf{q})$. In our case, this feature holds naturally for the composite bosons (the vortices are holes of physical fermions but not physical bosons). Thus we believe a good trial is replacing the left density operator of the composite bosons by
\begin{eqnarray} \label{eq:modified-density}
\rho_{B, L}(\mathbf{q})
\quad\longrightarrow\quad
\rho_{B, L}(\mathbf{q})-\rho_{B, R}(\mathbf{q})-\rho_{F, R}(\mathbf{q}).
\end{eqnarray}
It is very important that this modified density operator also satisfies the GMP algebra. It indicates that the modified density can be a legitimate density operator after the LLL projection. The correctness of such a trial can only be confirmed by a consistency check with the existing results, which will be performed later in Secs.~\ref{sec:single} and~\ref{sec:interaction}.

Based on the above discussion, we replace the Hamiltonian in Eq.~\eqref{eq:H0} with the operators in Eq.~\eqref{eq:modified-density}. The resulting modified Hamiltonian can be rearranged and separated in the following form,
\begin{eqnarray} \label{eq:H-modified}
H'=H_{BB}+H_{BF}+H_{FF}+H_F.
\end{eqnarray}
Each term is given explicitly by
\begin{widetext}
\begin{align}
H_{BB} \label{eq:HBB}
=~&\frac{1}{2}\int \frac{d^2\mathbf{q}}{(2\pi)^2} U_{ff}(\mathbf{q})
\left[\rho_{B,L}(\mathbf{q})-\rho_{B,R}(\mathbf{q})\right]
\left[\rho_{B,L}(-\mathbf{q})-\rho_{B,R}(-\mathbf{q})\right],
\\ \nonumber \\ \nonumber
H_{BF}
=~&\frac{1}{2}\int \frac{d^2\mathbf{q}}{(2\pi)^2}
\left[U_{bf}(\mathbf{q})\rho_{F,L}(\mathbf{q})
-U_{ff}(\mathbf{q})\rho_{F,R}(\mathbf{q})\right]
\left[\rho_{B,L}(-\mathbf{q})-\rho_{B,R}(-\mathbf{q})\right]
\\ \label{eq:HBF}
&+\frac{1}{2}\int \frac{d^2\mathbf{q}}{(2\pi)^2}
\left[\rho_{B,L}(\mathbf{q})-\rho_{B,R}(\mathbf{q})\right]
\left[U_{bf}(\mathbf{q})\rho_{F,L}(-\mathbf{q})
-U_{ff}(\mathbf{q})\rho_{F,R}(-\mathbf{q})\right],
\\ \nonumber \\ \nonumber
H_{FF}
=~&\frac{1}{2}\int \frac{d^2\mathbf{q}}{(2\pi)^2}
\left[U_{bb}(\mathbf{q})
:\rho_{F,L}(\mathbf{q})\rho_{F,L}(\mathbf{-q}):
-U_{bf}(\mathbf{q}):\rho_{F,L}(\mathbf{q})\rho_{F,R}(-\mathbf{q}):
\right]
\\  \label{eq:HFF}
&+\frac{1}{2}\int \frac{d^2\mathbf{q}}{(2\pi)^2}
\left[
U_{ff}(\mathbf{q}):\rho_{F,R}(\mathbf{q})\rho_{F,R}(\mathbf{-q}):
-U_{bf}(\mathbf{q}):\rho_{F,R}(\mathbf{q})\rho_{F,L}(-\mathbf{q}):
\right],
\\ \nonumber \\
H_F \label{eq:HF}
=~&\frac{1}{2}\int \frac{d^2\mathbf{q}}{(2\pi)^2}
\int\frac{d^2\mathbf{k}}{(2\pi)^2}
c_{\mathbf{k}}^\dagger c_{\mathbf{k}}
\left[
U_{ff}(\mathbf{q})
-U_{bf}(\mathbf{q})\left(e^{i\mathbf{k}\times\mathbf{q}\ell_B^2}
+e^{-i\mathbf{k}\times\mathbf{q}\ell_B^2}\right)
\right].
\end{align}
\end{widetext}
The physical meaning of each term is transparent. First, $H_{BB}$ describes the interaction between composite bosons which is inherited from the interaction between physical fermions. Next, $H_{BF}$ describes possible interaction between composite fermions and composite bosons after the replacement of density operators. The interaction between composite fermions is described by $H_{FF}$. Finally, $H_F$ is the \textit{one-body} term of composite fermion (the fermionic dipole formed by a physical boson and a hole of physical fermion).

\subsection{Exact solution for a single composite fermion}
\label{sec:single}

In the special case of having a single CF, the dimensions of Hilbert spaces for composite particles and physical particles match. This enables us to deduce the dispersion of a single composite fermion exactly from the Pasquier-Haldane construction. We obtain the dispersion by evaluating $\langle H_F\rangle$. Suppose the composite fermion has a momentum
$\mathbf{k}$. In addition, the $N-1$ composite bosons are condensed so that they do not play a role here. Then, one has
\begin{align} \label{eq:single-CF}
\nonumber
E(\mathbf{k})
&=\langle \mathbf{k}|H_F|\mathbf{k}\rangle
-\langle \mathbf{0}|H_F|\mathbf{0}\rangle
\\
&=\int \frac{d^2\mathbf{q}}{(2\pi)^2}
U_{bf}(\mathbf{q})
\left[1-\cos{\left(\mathbf{k}\times\mathbf{q}\ell_B^2\right)}\right].
\end{align}
The subtraction is essential since the dipole at zero momentum has a nonzero energy. When $U_{bf}(\mathbf{q})$ is isotropic, then the dispersion is also isotropic. In this scenario, we have
\begin{eqnarray} \label{eq:single-CF-iso}
E(k)
=\frac{1}{2\pi} \int_0^\infty q \tilde{u}_{bf}(q) e^{-q^2\ell_B^2/2}
\left[1-J_0(kq\ell_B^2)\right]dq.
\end{eqnarray}
The above result agrees with the exact single-exciton dispersion in Ref.~\cite{Yang2001}, which also applies to the CF dispersion here.

When $\nu_b\rightarrow 0$, the composite fermions have their effective masses dominated by the single-particle dispersion [Eq.~\eqref{eq:single-CF} or Eq.~\eqref{eq:single-CF-iso}]. The effective mass is determined by the dispersion at $|\mathbf{k}|\approx k_F$. In the case of an isotropic interaction, we have~\cite{Read1998, Senthil2020}:
\begin{align} \label{eq:effective mass}
\nonumber
\frac{1}{m_F^*}
&=\frac{1}{k_F}\left.\frac{\partial E(k)}{\partial k}\right|_{k=k_F}
\\
&=\frac{1}{2\pi k_F\ell_B}
\int_0^\infty s^2\tilde{u}_{bf}(s) e^{-s^2/2}J_1(k_F\ell_B s)~ds.
\end{align}
The dimensionless variable $s=q\ell_B$ has been defined.

The most promising platform in realizing the Bose-Fermi mixture will be cold atomic systems. At low temperature, the boson-boson and boson-fermion interactions are dominated by the $s$-wave scattering. In position space, the interaction takes the form of a Dirac-delta function. One has $u_{bf}(\mathbf{x})=u_{bf}\delta(\mathbf{x})$ and $\tilde{u}_{bf}(q)=u_{bf}$. The interaction strength $u_{bf}$ is determined by the $s$-wave scattering length of the particles. Note that both $u_{bb}$ and $u_{bf}$ can be tuned in the experiment by employing the technique of Feshbach resonance~\cite{Zaccanti2006, Bongs2006, Chin-RMP}. The single-exciton dispersion is given by
\begin{eqnarray}  \label{eq:single-CF-delta}
E_\delta(k)
=\frac{u_{bf}}{2\pi\ell_B^2}\left(1-e^{-k^2\ell_B^2/2}\right).
\end{eqnarray}
Up to a numerical factor, the result agrees with Ref.~\cite{Senthil2020}. However, the interaction here is characterized by $u_{bf}$ but not $u_{bb}$. This is consistent with the physical picture in Fig.~\ref{fig:dipole}. The corresponding inverse effective mass is
\begin{eqnarray} \label{eq:m-CF-delta}
\frac{1}{m_F^*}
=\left.\frac{1}{k_F}\frac{dE_{\delta}(k)}{dk}\right|_{k=k_F}
=\frac{u_{bf}}{2\pi}e^{-\nu_b}.
\end{eqnarray}
As $\nu_b\rightarrow 0$ with $N_f\rightarrow 1$, the result $m_F^*\rightarrow 2\pi/u_{bf}$ becomes asymptotically exact, because it receives no renormalization from CF-CF interaction and gauge fluctuations, for reasons we illustrate below. \\

\subsection{CF-CF scattering matrix element}
\label{sec:interaction}

Now, we discuss the interaction between composite fermions. Its most relevant term is given by the four-fermion interaction. Since the CFs in the mixture are spin-polarized, the interaction term written in the position space should involve two spatial derivatives. As a result, the interaction is irrelevant in the dilute limit under renormalization group (RG) transformation~\cite{Sachdev}. This means in the regime of $\nu_b\ll 1$, the effect from the CF-CF interaction is negligible. Therefore, the single-particle dispersion and effective mass in Sec.~\ref{sec:single} can accurately describe the properties of CFs in the low-density limit.

In spite of its irrelevance, it is important to check the scattering matrix element of the CF-CF interaction term for validating the modified Hamiltonian in Eq.~\eqref{eq:H-modified}. We denote the initial and final quantum states of the two CFs as 
$|\mathbf{p}_1, \mathbf{p}_2\rangle$ and $|\mathbf{k}_1, \mathbf{k}_2\rangle$. In the discussion below, we only consider the regime of small momentum by focusing on the leading order terms.

Since $H_F$ in Eq.~\eqref{eq:HF} is a one-body term, we should not include it in the two-body interaction. For $H_{BB}$ in Eq.~\eqref{eq:HBB}, it leads to a contribution in the order of
$|\mathbf{k}|^2$. Considering the second line of $H_{BF}$ in Eq.~\eqref{eq:HBF}, one obtains
\begin{widetext}
\begin{align}
\nonumber
&\langle \mathbf{p}_1, \mathbf{p}_2, \text{CB}
\left| \left[\rho_{B,L}(\mathbf{q})-\rho_{B,R}(\mathbf{q})\right]
\left[U_{bf}(\mathbf{q})\rho_{F,L}(-\mathbf{q})
-U_{ff}(\mathbf{q})\rho_{F,R}(-\mathbf{q})\right]
\right|
\mathbf{k}_1, \mathbf{k}_2, \text{CB}\rangle
\\
\sim
~&
\langle \mathbf{p}_1, \mathbf{p}_2
\left| c_{\mathbf{k}+\mathbf{q}}^\dagger c_{\mathbf{k}}\right|
\mathbf{k}_1, \mathbf{k}_2\rangle
\langle\text{CB}\left|b_{\mathbf{k}'-\mathbf{q}}b_{\mathbf{k}'}\right|\text{CB}\rangle
\sin{\left(\frac{\mathbf{k'}\times\mathbf{q}}{2}\ell_B^2\right)}.
\end{align}
\end{widetext}
Here, $|\text{CB}\rangle$ denotes the quantum state of the $N-2$ composite bosons which is separated from the two composite fermions. The above matrix element vanishes as $\langle\text{CB}\left|b_{\mathbf{k}'-\mathbf{q}}b_{\mathbf{k}'}\right|\text{CB}\rangle$ requires $\mathbf{q}=0$. The same conclusion is reached for another term in $H_{BF}$. Therefore, in the leading order of momentum, we are left with a two-body interaction term $V=H_{FF}$. From this, we have
\begin{widetext}
\begin{align}
\nonumber
\langle \mathbf{p}_1, \mathbf{p}_2\left| V\right|\mathbf{k}_1, \mathbf{k}_2\rangle
\approx~&\frac{1}{2}\int\frac{d^2\mathbf{q}}{(2\pi)^2}
\int\frac{d^2\mathbf{k} d^2\mathbf{k'}}{(2\pi)^4}
\langle \mathbf{p}_1, \mathbf{p}_2
| c_{\mathbf{k}-\mathbf{q}}^\dagger c_{\mathbf{k'}+\mathbf{q}}^\dagger
c_{\mathbf{k'}} c_{\mathbf{k}} |
\mathbf{k}_1, \mathbf{k}_2\rangle
\left[U_{bb}(\mathbf{q})-U_{bf}(\mathbf{q})
+U_{ff}(\mathbf{q})-U_{bf}(\mathbf{q})
\right]
\\
=~&\frac{\delta(\mathbf{k}_1+\mathbf{k}_2-\mathbf{p}_1-\mathbf{p}_2)}{A}
\left[U_{bb}(\mathbf{k}_1-\mathbf{p}_1)-U_{bf}(\mathbf{k}_1-\mathbf{p}_1)
+U_{ff}(\mathbf{k}_1-\mathbf{p}_1)-U_{bf}(\mathbf{k}_1-\mathbf{p}_1)
\right].
\end{align}
\end{widetext}
Here, the symbol $A$ denotes the area of the system. The matrix element is consistent with the effective dipole-dipole interaction term in Ref.~\cite{Lian-Zhang}. This is more evidence to support the replacement of the density operator and the modified Hamiltonian in Sec.~\ref{sec:pre-H}.

\section{Hartree-Fock approximation of CF-CF interaction}
\label{sec:HF-approximation}

To obtain the correction in the effective mass, one needs to include the effect from the two-body interaction. A simple treatment is the Hartree-Fock approximation, which will be employed in the following discussion.

For the Bose-Fermi mixture, we assume the wave function of the whole system is factorized into two parts. Specifically, the composite fermions takes a wave function in the Slater determinant form, whereas the composite bosons have a wave function in a permanent form. Then, we have the following properties for the contractions:
\begin{align}
\langle c^\dagger_{\mathbf{k}}c_{\mathbf{k}'}\rangle
=(2\pi)^2\delta(\mathbf{k}-\mathbf{k}')n_F(\mathbf{k}),
\\
\langle b^\dagger_{\mathbf{k}}b_{\mathbf{k}'}\rangle
=(2\pi)^2\delta(\mathbf{k}-\mathbf{k}')n_B(\mathbf{k}).
\end{align}
Here, $n_F(\mathbf{k})$ labels the occupation number of composite fermions at momentum
$\mathbf{k}$. Note that the prefactor $(2\pi)^2$ originates from the commutation and anticommutation relations in Eqs~\eqref{eq:anticommutator} and~\eqref{eq:commutator}. From the contractions, $H_{BF}$ gives zero contribution to the energy expectation value. By including both the single-particle dispersion and the Hartree-Fock result, we obtain the energy dispersion for composite fermions:
\begin{widetext}
\begin{align} \label{eq:dispersion-CF}
\nonumber
\epsilon_F(\mathbf{k})
=&~\int \frac{d^2\mathbf{q}}{(2\pi)^2}
U_{bf}(\mathbf{q})
\left[1-\cos{\left(\mathbf{k}\times\mathbf{q}\ell_B^2\right)}
\right]
\\
&-\int\frac{d^2\mathbf{k'}}{(2\pi)^2}
\left[U_{bb}(\mathbf{k}-\mathbf{k'})+U_{ff}(\mathbf{k}-\mathbf{k'})
-2U_{bf}(\mathbf{k}-\mathbf{k'})\cos{\left(\mathbf{k}\times\mathbf{k'}\ell_B^2\right)}
\right]n_F(\mathbf{k'}).
\end{align}
\end{widetext}

In the special case when all interactions are isotropic, the composite fermions will form a circular Fermi surface with the Fermi momentum $k_F=\sqrt{2\nu_b}\ell_B^{-1}$. Then, Eq.~\eqref{eq:dispersion-CF} can be simplified in to
\begin{widetext}
\begin{align}
\nonumber
\epsilon_F(k)
=&~\frac{1}{2\pi}\int q \tilde{u}_{bf}(q) e^{-q^2\ell_B^2/2}
\left[1-J_0(kq\ell_B^2)\right]~dq
\\
&-e^{-k^2\ell_B^2/2}\int_0^{k_F}\frac{dk'}{2\pi}
k' e^{-k'^2\ell_B^2/2}
\left\{
\left[U_{bb}(|\mathbf{k}-\mathbf{k'}|)+U_{ff}(|\mathbf{k}-\mathbf{k'}|)\right]
I_0(kk'\ell_B^2)
-2U_{bf}(|\mathbf{k}-\mathbf{k'}|)
\right\}.
\end{align}
\end{widetext}
The second term can be expanded in a power series of $k_F\ell_B\sim\sqrt{\nu_b}$. These resulting terms become irrelevant in the dilute limit, i.e. $\nu_b\rightarrow 0$.

For composite bosons, the dispersion and effective mass are solely contributed by
$H_{BB}$ in Eq.~\eqref{eq:HBB}. The Hartree-Fock approximation leads to the dispersion,
\begin{widetext}
\begin{align} \label{eq:dispersion-CB}
\epsilon_B(\mathbf{k})
=\int \frac{d^2\mathbf{q}}{(2\pi)^2}
U_{ff}(\mathbf{q})
\left[1-\cos{\left(\mathbf{k}\times\mathbf{q}\ell_B^2\right)}
\right]
+2\int\frac{d^2\mathbf{k'}}{(2\pi)^2}
U_{ff}(\mathbf{k}-\mathbf{k'})
\left[1-\cos{\left(\mathbf{k}\times\mathbf{k'}\ell_B^2\right)}\right]n_B(\mathbf{k'}).
\end{align}
\end{widetext}
In cold atomic systems, the most dominant interaction between spin-polarized fermions is the $p$-wave scattering, which is irrelevant in the RG sense. Independent of the details of the fermion-fermion interaction, the non-negative $\epsilon_B(\mathbf{k})$ favors a condensation of composite bosons in the single-particle state with $\mathbf{k}=\mathbf{0}$.

\section{Non-commutative field theory and gauge fluctuations}
\label{sec:gauge}

Using the effective masses of CFs and CBs, the Hamiltonian describing the Bose-Fermi mixture under the Hartree-Fock approximation can be written as
\begin{eqnarray} \label{eq:H-HF}
H_{\rm HF}
=\int\frac{d^2\mathbf{k}}{(2\pi)^2}
\left(
\frac{k^2}{2m_F^*}c^\dagger_\mathbf{k}c_\mathbf{k}
+\frac{k^2}{2m_B^*}b^\dagger_\mathbf{k}b_\mathbf{k}
\right).
\end{eqnarray}
Here, the approximation of the energy dispersion by a simple quadratic form is justified as the low-energy physics of the system is dominated by quantum states that are close to the Fermi surface. Furthermore, it is noted that an exact dispersion of composite fermions in the low-density limit was obtained in Sec.~\ref{sec:single}. In this limit, the Fermi surface is small and close to the origin, so that the quadratic dispersion is really an expansion from the origin. Note that $H_{\rm HF}$ does not contain any explicit term for the CF-CB interaction. The interaction has been transmuted into different terms in $H_{FF}$ and $H_F$ after employing the modified density operators in Eq.~\eqref{eq:modified-density}. Furthermore, its residual effect described by $H_{BF}$ gives no contribution under the Hartree-Fock approximation. The corresponding Euclidean action for $H_{\rm HF}$ is
\begin{align} \label{eq:S-HF}
\nonumber
S_{\rm HF}
=&\int\frac{d^3k}{(2\pi)^2}
\left[
\bar{c}(\mathbf{k},\tau)\frac{\partial c(\mathbf{k},\tau)}{\partial\tau}
+\frac{k^2}{2m_F^*}\bar{c}(\mathbf{k},\tau)c(\mathbf{k},\tau)\right]
\\
&+\int\frac{d^3k}{(2\pi)^2}
\left[\bar{b}(\mathbf{k},\tau)\frac{\partial b(\mathbf{k},\tau)}{\partial\tau}
+\frac{k^2}{2m_B^*}\bar{b}(\mathbf{k},\tau)b(\mathbf{k},\tau)
\right].
\end{align}
Here, $\tau$ labels the imaginary time. To save space, we have defined the shorthand notation, $d^3k=d^2\mathbf{k}d\tau$.

Equations~\eqref{eq:H-HF} and~\eqref{eq:S-HF} seem to suggest that the Hartree-Fock theory of CFs and CBs look like those for particles living in the ordinary commutative space, whose range of momenta is infinite even if the system size is finite. On the other hand, the number of independent momenta is $N^2$ in our case, reflecting the finite number of degrees of freedom in the LLL. The issue becomes clear when we remind ourselves that particles restricted to the LLL have their dynamics controlled by the noncommutative guiding center coordinates $\mathbf{R}$. This feature suggests us to formulate our theory in noncommutative space, instead of ordinary commutative space. Furthermore, we pointed out in Sec.~\ref{sec:mixture} the existence of a SU$(N)_R\times$ U$(1)$ gauge invariance under the transformation of Landau orbitals for vortices. This is reinstated by the constraint in Eq.~\eqref{eq:constraint}. In order to implement the constraint, we need to introduce gauge fields, such that the physical Hilbert space is made of the gauge-invariant states. This in turn introduces gauge interactions among the CFs and CBs, which plays a crucial role in the physics. To implement these, we follow the procedures in Ref.~\cite{Senthil2020} to construct a low-energy field theory of the mixture. Meanwhile, the striking difference between CF liquids in the Bose-Fermi mixture and system of bosons at $\nu=1$ will be addressed.

From the above discussion, it becomes quite natural to study gauge fluctuation by defining the CF field as a function of $\mathbf{R}$:
\begin{align}
c(\mathbf{R},\tau)
&=\int\frac{d^2\mathbf{k}}{(2\pi)^{3/2}}\exp{(i\mathbf{k}\cdot\mathbf{R})}~
c(\mathbf{k},\tau),
\\
\bar{c}(\mathbf{R},\tau)
&=\int\frac{d^2\mathbf{k}}{(2\pi)^{3/2}}\exp{(-i\mathbf{k}\cdot\mathbf{R})}~
\bar{c}(\mathbf{k},\tau).
\end{align}
Similar definitions hold for composite boson fields. This approach was employed By Dong and Senthil. It is possible to trade the noncommutative fields as fields defined on the commutative spacetime $(\mathbf{x}, \tau)$. Nevertheless, the corresponding fields $c(\mathbf{x},\tau)$ and $\bar{c}(\mathbf{x},\tau)$ are still noncommutative fields. This is captured by replacing the multiplication between ordinary fields with the Moyal-Weyl star product $\star$~\cite{Douglas-RMP, Szabo}. In the present case, we have~\cite{Milovanovic2021}
\begin{align}
c(\mathbf{x})\star
c(\mathbf{y})
=\exp{\left(-i\ell_B^2\epsilon^{ij}
\frac{\partial}{\partial x^i}\frac{\partial}{\partial y^j}\right)}
\lim_{\mathbf{x}\rightarrow\mathbf{y}}
c(\mathbf{x})c(\mathbf{y}).
\end{align}
Mathematically, the star product provides a deformed multiplication law that implements the noncommutative structure of the guiding center coordinates. Using the noncommutative fields, the action of the Bose-Fermi mixture can be written as
\begin{align}
\nonumber
&S_{\rm HF}
\\ \nonumber
=&\int
\left[
\bar{c}(\mathbf{x},\tau) \star \partial_\tau c(\mathbf{x}, \tau)
+\frac{\nabla\bar{c}(\mathbf{x},\tau)\star\nabla c(\mathbf{x},\tau)}{2m_F^*}
\right]d^3x
\\
&+\int
\left[
\bar{b}(\mathbf{x},\tau) \star \partial_\tau b(\mathbf{x}, \tau)
+\frac{\nabla\bar{b}(\mathbf{x},\tau)\star\nabla b(\mathbf{x},\tau)}{2m_B^*}
\right]d^3x.
\end{align}
Here, the symbol $d^3x=d^2\mathbf{x}d\tau$ is defined.

Now, we discuss the coupling between composite particles and (emergent) gauge fields. When
$\mathbf{q}\neq\mathbf{0}$, $c_{\mathbf{k}}^\dagger c_{\mathbf{k}}$ and
$b_{\mathbf{k}}^\dagger b_{\mathbf{k}}$  in Eq.~\eqref{eq:H-HF} do not commute with
$\rho_{F,R}(\mathbf{q})$ and $\rho_{B,R}(\mathbf{q})$, respectively. It implies the SU$(N)_R$ (gauge) symmetry is spontaneously broken in the Hartree-Fock ground state. As a result the corresponding SU$(N)$ gauge fluctuation is Higgsed and becomes unimportant at low-energy, just like in the system of bosons at $\nu=1$~\cite{Senthil2020}. In that case the remaining U(1) symmetry is respected by the CF Fermi sea ground state, 
As a result it was necessary to introduce a noncommutative emergent U$(1)$ gauge field $a$, which couples to the right current densities of the composite fermions. Its associated gauge fluctuation in the small momentum $\mathbf{q}\approx \mathbf{0}$ (or long-wavelength) regime becomes the most important gauge fluctuation there. Here we follow the same procedure, but as we will see later in Sec.~\ref{sec:Higgs}, the existence of composite boson condensate generates a mass gap to Higgs the U$(1)$ gauge fluctuation as well, with the Higgs mass depending sensitively on $\nu_f$ or CB density, vanishing for $\nu_f=0$ (or equivalently, $\nu_b=1$, which is the case studied in Ref.~\cite{Senthil2020}). This feature does not exist in the system of bosons at $\nu=1$. Furthermore, we introduce a noncommutative background gauge field $A$ and couple it to the left current densities of the composite particles. Its physical meaning will become more transparent when we approximate $A$ by its commutative field later. For a more detailed discussion on the gauge transformation, we refer the readers to Ref.~\cite{Senthil2020}.

Following the above discussion, we define the covariant derivatives for the CF and CB fields as~\cite{Senthil2020}
\begin{align}
D_\mu c=\partial_\mu c-ic\star a_\mu-i A_\mu \star c,
\\
D_\mu b=\partial_\mu b-ib\star a_\mu-i A_\mu \star b.
\end{align}
When there is no confusion, we hide the argument of $c$ and $b$. They are functions of
$\mathbf{x}$ and  $\tau$. Using the covariant derivatives, a gauge invariant action can be formulated:
\begin{widetext}
\begin{eqnarray} \label{eq:S}
S
=\int d^2\mathbf{x}d\tau~
\left[\bar{c}\star D_\tau c+\sum_{j=1}^2\frac{1}{2m_F^*}\overline{D}_j c\star D_j c
+\bar{b}\star D_\tau b+\sum_{j=1}^2\frac{1}{2m_B^*}\overline{D}_j b\star D_j b
+ia_0\bar{\rho}\right].
\end{eqnarray}
\end{widetext}
Here, an additional term $ia_0\bar{\rho}$ is included in the action. Its significance becomes transparent if we consider the equation of motion for $a_0$. This leads to
\begin{eqnarray}
\bar{c}\star c+\bar{b}\star b
=\bar{\rho}
=\frac{1}{2\pi\ell_B^2}.
\end{eqnarray}
The equation reproduces the constraint in Eq.~\eqref{eq:constraint} and fixes the average number density of composite particles at $\bar{\rho}=1/2\pi\ell_B^2$.

\subsection{The Seiberg-Witten map}

Since the action $S$ is a noncommutative field theory, it is difficult to reveal the low-energy physics of the Bose-Fermi mixture. On the other hand, the Seiberg-Witten map provides a systematic approach to express the noncommutative field in a power series of the noncommutative parameter~\cite{Seiberg-Witten}. Each term in the power series consists of the commutative field, gauge fields, and their derivatives. In describing the physics of composite fermions within the LLL, the noncommutative parameter is $\Theta=-\ell_B^2$. Using the map, the noncommutative composite particle fields and U$(1)$ gauge fields are approximated as~\cite{Senthil2020}:
\begin{align} 
\label{eq:SW-A}
A_\mu
&=\hat{A}_\mu-\frac{\Theta}{2}\epsilon^{\alpha\beta}\hat{A}_\alpha
\left(\partial_\beta\hat{A}_\mu+\partial_\beta\hat{A}_\mu-\partial_\mu\hat{A}_\beta
\right),
\\
\label{eq:SW-a}
a_\mu
&=\hat{a}_\mu+\frac{\Theta}{2}\epsilon^{\alpha\beta}\hat{a}_\alpha
\left(\partial_\beta\hat{a}_\mu+\partial_\beta\hat{a}_\mu-\partial_\mu\hat{a}_\beta
\right),
\\
\label{eq:SW-c}
c
&=\psi+\frac{\Theta}{2}\epsilon^{\alpha\beta}
\left[(\hat{a}_\alpha-\hat{A}_\alpha)\partial_\beta\psi
-i\hat{a}_\alpha\hat{A}_\beta\psi\right],
\\
\label{eq:SW-b}
b
&=\phi+\frac{\Theta}{2}\epsilon^{\alpha\beta}
\left[(\hat{a}_\alpha-\hat{A}_\alpha)\partial_\beta\phi
-i\hat{a}_\alpha\hat{A}_\beta\phi\right].
\end{align}
Since our goal is to obtain an approximate low-energy commutative field theory in the leading order of $\Theta$, the series is truncated at the first power of $\Theta$. It seems that the validity of the expansion relies on the smallness of $|\Theta|$. However, this makes no sense as $\ell_B^2$ is a dimensionful quantity. This issue and the corresponding assumptions in applying the Seiberg-Witten map will be discussed in Sec.~\ref{sec:validity}. Note that we follow the convention in recent literature, such that $\hat{A}$, $\hat{a}$, $\psi$, and $\phi$ are commutative fields. Being commutative fields, any product between them is defined by the ordinary multiplication law.

In the zeroth order in $\Theta$, the approximate Lagrangian density is given by
\begin{align}
\nonumber
\mathcal{L}_0
=~&\bar{\psi}(\partial_\tau-i \hat{a}_0-i \hat{A}_0)\psi
+\sum_{j=1}^2
\frac{|(\partial_j -i \hat{a}_j-i \hat{A}_j)\psi|^2}{2m_F^*}
\\ \nonumber
&+\bar{\phi}(\partial_\tau-i \hat{a}_0-i \hat{A}_0)\phi
+\sum_{j=1}^2
\frac{|(\partial_j -i \hat{a}_j-i \hat{A}_j)\phi|^2}{2m_B^*}
\\
&+i\left[\hat{a}_0+\frac{\Theta}{2}\epsilon^{ij}\hat{a}_i
\left(\partial_j\hat{a}_0+\partial_j\hat{a}_0-\partial_0\hat{a}_j
\right)\right]\bar{\rho}.
\end{align}
Basically, one simply replaces all noncommutative fields by their corresponding commutative fields. Now, $\partial_1\hat{A}_2-\partial_2\hat{A}_1$ should be interpreted as the additional magnetic field different from the one which defines the filling factor of the system. It is important to treat the term $ia_0\bar{\rho}$ in Eq.~\eqref{eq:S} carefully. Since $\bar{\rho}\sim \ell_B^{-2}\sim 1/|\Theta|$, one should also include the first order correction term of $a_0$ in the approximation. After performing an integration by parts in the last line of $\mathcal{L}_0$, one obtains
\begin{widetext}
\begin{align}
\nonumber
\mathcal{L}_0
=~&\bar{\psi}[\partial_\tau-i (\hat{a}_0+ \hat{A}_0)]\psi
+\frac{1}{2m_F^*}\sum_{j=1}^2
\left|[\partial_j -i (\hat{a}_j+ \hat{A}_j)]\psi \right|^2
+i\hat{a}_0\bar{\rho}
-\frac{i}{4\pi}\epsilon^{\alpha\beta\gamma}a_\alpha\partial_\beta a_\gamma
\\
&+\bar{\phi}[\partial_\tau-i (\hat{a}_0+ \hat{A}_0)]\phi
+\frac{1}{2m_B^*}\sum_{j=1}^2
|[\partial_j -i (\hat{a}_j+ \hat{A}_j)]\phi|^2.
\end{align}
\end{widetext}
The first line above resembles the form of HLR theory~\cite{HLR}. As a result, the composite fermions in the mixture will form a composite Fermi liquid. However, the underlying physics is quite different. Here, the effective mass for the composite fermions is solely contributed by the interaction between physical particles. It exactly approaches Eq.~\eqref{eq:effective mass} in the limit of $\nu_b\rightarrow 0$. More importantly, the CFs here are defined within the LLL which is not the case in the original HLR theory. The Chern-Simons term has the correct coefficient $1/4\pi$ for a system with filling factor $\nu=1$. This is correctly reproduced only if the noncommutative gauge field $a_0$ is coupled identically to both composite fermions and composite bosons in the mixture.

It is natural to ask for the first-order correction terms in $\Theta$ in the Lagrangian density. The calculation turns out to be cumbersome. On the other hand, a comparison between the Seiberg-Witten map in Eqs.~\eqref{eq:SW-A}-\eqref{eq:SW-b} and the map in Ref.~\cite{Senthil2020} suggests that they take a very similar form. The main difference is that there is an additional composite boson field $b$ in the present system. In principle, one can follow the detailed calculation in Ref.~\cite{Senthil2020} and obtain the $O(\Theta)$ terms in the Lagrangian density of the Bose-Fermi mixture. We expect there will be a correction term describing the coupling between internal electric field and density gradient of the composite particles, similar to the system of bosons at $\nu=1$ as shown in the reference. In addition, a correction term that goes as $\delta \rho_L \ell_B^2$ will present. As we will show below, both correction terms in $O(\Theta)$ are small in the low-density regime of composite fermions.

\subsection{Validity of the Seiberg-Witten map}
\label{sec:validity}

Now, we should comment on the validity of the application of Seiberg-Witten map in the Bose-Fermi mixture. It is obvious that the parameter $|\Theta|=\ell_B^2$ in the expansion is not a dimensionless parameter. As pointed out precisely in Ref.~\cite{Senthil2020}, the small parameters that one should consider are $\delta\rho_L|\Theta|$ and $q^2|\Theta|$. Here, $\delta\rho_L$ stands for the deviation in the density of composite particles away from their mean values in the real space. The momentum of the gauge field is denoted as $q$.

In the present case, the density fluctuation of the composite bosons in real space is believed to be small due to the Bose-Einstein condensation. Hence, $\delta\rho_L$ is governed by the density fluctuation of composite fermions. This fluctuation is bounded by the filling factor of physical bosons, $\nu_b=N_b/N$. In the dilute limit where $\nu_b\ll 1$, $\delta\rho_L\ell_B^2$ is a small parameter by default. Furthermore, the typical momentum of the gauge field is set by the Fermi momentum, i.e. $q\sim k_F$. Since $k_F^2\ell_B^2\sim\nu_b$, we also have $q^2|\Theta|$ being a small parameter when $\nu_b\ll 1$. Therefore, the application of Seiberg-Witten map is manifestly justified when the mixture has a small number of bosons. In principle, this regime can be accessed experimentally. This makes it advantageous to study the Bose-Fermi mixture as compared to other systems. 

The zeroth order terms in the Lagrangian density captures the most important and interesting physics of the system, which we will discuss below.

\subsection{Anderson-Higgs mass of gauge fields due to composite boson condensation}
\label{sec:Higgs}

As mentioned in the introduction, the existence of composite bosons and their condensation lead to a dramatic difference between the Bose-Fermi mixture and the system of bosons at $\nu=1$. Due to the CB condensation, we may write the composite boson field as
\begin{eqnarray} \label{eq:BEC-trial}
\phi(x,\tau)=\sqrt{n_B}e^{i\eta(x,\tau)}.
\end{eqnarray}
Here, $n_B$ denotes the average number density of bosons in the condensate. In the mean-field approximation, $n_B$ is assumed to be uniform and independent of position. For $\eta(x,\tau)$, it is a slow varying phase factor. It can be gauged away (or ``eaten by the gauge boson") by making a transformation in the electromagnetic potential:
\begin{eqnarray} \label{eq:gauge-phase}
\hat{a}_\mu\rightarrow \hat{a}_\mu+\partial_\mu\eta.
\end{eqnarray}
Since the Bose-Fermi mixture is a two dimensional system, it seems that the formation of BEC of composite bosons violates the Mermin-Wagner theorem~\cite{MW1966, Hohenberg}. However, the composite bosons are coupled to the emergent gauge field. Due to this gauge coupling, the condensate wave function (or order parameter) is a gauge-dependent quantity, hence not a physical observable. In other words, a nonvanishing vacuum expectation value for $\phi(x,\tau)$ does not imply a spontaneous symmetry breaking. In fact, the BEC of composite bosons does not possess a true long-range order. 

Let us focus on the Hamiltonian of the composite bosons. For simplicity, we set the additional external gauge potential $\hat{A}=0$. Then, a direct substitution of Eqs.~\eqref{eq:BEC-trial} and~\eqref{eq:gauge-phase} leads to a quadratic term for the emergent gauge field:
\begin{eqnarray}
H_{CB}
=\frac{1}{2m_B^*}\sum_{j=1}^2
|\left(\partial_j -i \hat{a}_j\right)\phi|^2
=\frac{n_B}{2m_B^*}|\hat{a}|^2.
\end{eqnarray}
The gauge field acquires a mass gap, and the corresponding gauge boson has a mass
\begin{eqnarray} \label{eq:gauge-mass}
M
=\sqrt{\frac{n_B}{m_B^*}}
=\sqrt{\frac{1-\nu_b}{2\pi\ell_B^2 m_B^*}}.
\end{eqnarray}
This significantly changes the low-energy properties of composite fermions in the mixture, which will be discussed in the next subsection. Notice that the above discussion is similar to the Anderson-Higgs mechanism of composite bosons in explaining fractional quantum Hall effect by Zhang, Hansson, and Kivelson~\cite{ZHK1989}.

\subsection{Landau-Fermi liquid formed by CFs and crossover to non-Fermi liquid behavior}
\label{sec:crossover}

Now, we discuss the low-energy properties of composite fermions in the Bose-Fermi mixture. We assume the physical bosons and fermions in the mixture interact with a Dirac-delta type interaction, which is appropriate for cold atomic systems. In the mean-field approximation, the CFs form a circular Fermi surface with a Fermi momentum, $k_F=\sqrt{2\nu_b}/\ell_B$. The corresponding Fermi energy is given by
\begin{eqnarray} \label{eq:E-F}
E_F
=\frac{k_F^2}{2m_F^*}
\approx u_{bf}\left(\frac{\nu_b}{2\pi \ell_B^2}\right)e^{-\nu_b}.
\end{eqnarray}
Note that we have employed the result for effective mass in Eq.~\eqref{eq:m-CF-delta}, which is exact as $\nu_b\rightarrow 0$. Suppose $\ell_B$ is fixed. In the regime of $\nu_b\ll 1$, both $k_F$ and $E_F$ are very small. Meanwhile, the emergent gauge field acquires a mass gap with an associated energy scale,
\begin{eqnarray} \label{eq:E-M}
E_M=M^2=\frac{1-\nu_b}{2\pi\ell_B^2 m_B^*}.
\end{eqnarray}
Since $m_B^*$ depends on $\nu_b$, $E_M$ is actually a nonlinear function in $\nu_b$. One expects $E_M$ would grow faster as $\nu_f=1-\nu_b$ increases. At low temperature $(T)$ such that $k_B T\ll E_M$, the gauge field fluctuation is strongly suppressed. As stated in Sec.~\ref{sec:Higgs}, the Higgs mechanism does not involve spontaneous symmetry breaking, so the suppression is allowed even in two dimensions. The composite Fermi liquid behaves as a genuine Landau-Fermi liquid when $k_B T<E_F$. In this situation, we predict the system has a low-temperature specific heat,
\begin{eqnarray}
c_v=\frac{\pi}{6}m_F^* k_B^2 T
=\left(\frac{\pi^2 k_B^2}{3u_{bf}}e^{\nu_b}\right)T,
\end{eqnarray}
which can be measured in experiment.

In the opposite limit with $\nu_b\rightarrow 1$, $E_M\rightarrow 0$ and the gauge boson becomes massless. Also, both $k_F$ and $E_F$ are not small. When $E_M<k_B T<E_F$, fluctuation of gauge field becomes significant. The composite Fermi liquid becomes a non-Fermi liquid. In this scenario, the low-temperature specific heat is mainly contributed by gauge field fluctuation, and should scale as $T^{2/3}$ predicted in the HLR theory~\cite{HLR, footnote2}. It is noted that this $T^{2/3}$ dependence was actually obtained from the random phase approximation (RPA), which becomes exact only in the large-$N$ limit. Given that the temperature dependence from the leading order terms is already an approximate result from RPA, it is not worthwhile to discuss the correction in specific heat from higher-order terms in the Lagrangian density of the system. On the other hand, the linear $T$ dependence in the low-density regime is an exact result. Furthermore, the smallness of $\delta\rho_L|\Theta|$ and $q^2|\Theta|$ (see Sec.~\ref{sec:validity}) ensures that the correction in $c_v$ should be insignificant in the low-density regime.

From the above discussion, we predict there should be a crossover from Landau-Fermi liquid to non-Fermi liquid for composite fermions. The crossover occurs because $E_F$ and $E_M$ become comparable to each other when $\nu_b$ is increased from nearly zero to one. This idea has been summarized in Fig.~\ref{fig:crossover}. A smoking gun of the crossover would be an observation of the change in temperature dependence of the low-$T$ specific heat from $T$ to $T^{2/3}$ (or more qualitatively, a change in the temperature dependence in $c_v$). Such an observation will also provide a direct evidence of the existence of emergent gauge field in the system. 

\section{Summary and conclusion}

To conclude, we have formulated a low-energy description of composite fermions in the Bose-Fermi mixture at total Landau-level filling factor $\nu=1$. In order to restrict the theory in the lowest Landau level, we have generalized and employed the Pasquier-Haldane construction. Besides composite fermions, the construction involves composite bosons in the enlarged Hilbert space. The composite fermions have been identified as dipolar excitons, which form between bosons and fermionic holes in the mixture. Armed with this intuitive physical picture, we derived a preferred Hamiltonian to describe the system. Its validity was confirmed by evaluating the single-particle dispersion and two-body interaction matrix element for composite fermions. Both results turned out to be consistent with previous work.

Different from previous work, the feasibility of varying the CF density by changing the filling factors of the physical bosons and fermions, allowed us to examine the limit of having a small number of composite fermions. Since the four-fermion interaction is irrelevant in this limit, we showed that the energy dispersion and effective mass of CFs deduced from the preferred Hamiltonian are asymptotically exact as $\nu_b\rightarrow 0$. We applied the Hartree-Fock approximation to obtain the correction in energy dispersion due to the irrelevant CF-CF interaction.

We followed the recent work, and employed the techniques of noncommutative field theory and the Seiberg-Witten map to study the gauge fluctuation in the system. Very importantly, we pointed out the generation of a mass gap for the gauge field due to the CB condensation. This mechanism strongly suppresses the effect of gauge fluctuation at low temperature when the system has $\nu_b\ll 1$. On the other hand, the gauge field becomes massless and its fluctuation becomes significant when $\nu_b\rightarrow 1$. Therefore, we have proposed a crossover from Landau-Fermi liquid to non-Fermi liquid of composite fermions by increasing temperature. This result is illustrated in Fig.~\ref{fig:crossover}. We suggested a possible detection of the crossover by observing a change from $T$ to $T^{2/3}$-dependence in the low-temperature specific heat of the system. Furthermore, we have validated the application of the Seiberg-Witten map in the system by arguing the parameters, $\delta\rho_L\ell_B^2\ll1 $ and $q^2\ell_B^2\ll1$ are indeed small when $\nu_b\ll 1$.

Finally, let us recapitulate the advantages of studying the Bose-Fermi mixture at $\nu=1$. The tunability of CF density permitted us to (i) perform an asymptotically exact calculation of the effective mass of composite fermions; (ii) justify the application of Seiberg-Witten map in a  natural manner; and (iii) suggest a possible crossover between Landau-Fermi liquid and non-Fermi liquid of CFs in the system. To the best of our knowledge, no existing proposal can realize all these three features in a single system. Therefore, we have proposed an accessible system, in which a well-controlled and quantitative theory of composite fermions can be formulated properly within the lowest Landau level. We believe our work has provided an important and deeper understanding of composite fermions and the emergence of gauge fields in condensed matter systems.

\section*{Acknowledgments}

This research was supported by the National Science Foundation Grant No. DMR-1932796, and performed at the National High Magnetic Field Laboratory, which is supported by National Science Foundation Cooperative Agreement No. DMR-1644779, and the State of Florida.

\appendix


\begin{thebibliography}{99}
\bibitem{footnote1}
In general, one may attach $2n$ magnetic flux quanta to each electron. 
The theory of composite fermions predicts a gapless state for electrons at $\nu=1/2n$. 
Similarly, the attachment of $2n+1$ flux quanta to each boson also turns them into CFs, which we will use later in this paper.
\bibitem{footnote2}
If the electrons or particles interact with long-range Coulomb interaction, then the system will be a marginal Fermi liquid. The low-temperature specific heat contributed by gauge field fluctuation scales as $T\ln{T}$.
\end{thebibliography}
\end{document}